\documentclass [12pt,preprint]{aastex}
\usepackage{epsfig}
\begin{document}
\voffset-1cm
\newcommand{\gsim}{\hbox{\rlap{$^>$}$_\sim$}}
\newcommand{\lsim}{\hbox{\rlap{$^<$}$_\sim$}}

\title{Implications of the Large Polarization \\
Measured in Gamma Ray 
Bursts}

\author{Shlomo Dado\altaffilmark{1}, Arnon Dar\altaffilmark{1}
and A. De
R\'ujula\altaffilmark{2}}

\altaffiltext{1}{dado@phep3.technion.ac.il, arnon@physics.technion.ac.il,
dar@cern.ch.\\
Physics Department and Space Research Institute, Technion, Haifa 32000,
Israel}
\altaffiltext{2}{alvaro.derujula@cern.ch; Theory Unit, CERN,
1211 Geneva 23, Switzerland \\
Physics Department, Boston University, USA}

\begin{abstract}
The polarization of the prompt $\gamma$-ray emission has been 
measured in four bright gamma ray bursts (GRBs). It 
was nearly maximal in all cases, as predicted by the 
Cannonball (CB) model of GRBs long before the observations. 
These results are inconsistent with standard models of GRBs wherein the 
prompt emission is due to synchrotron radiation.
A much smaller linear polarization
is predicted by the CB model for the prompt emission in X-ray flashes 
(XRFs) and in extremely luminous GRBs.
These measurements would provide yet another stringent test of the 
CB model and its unification of GRBs and XRFs.
\end{abstract}

\keywords{gamma rays: bursts---polarization: general}

%\maketitle

\section{Introduction}

Polarization measurements of radiations from astronomical sources are an 
important diagnostic tool of their means of production. Gamma ray bursts 
(GRBs) are not an exception. The polarization of the prompt $\gamma$-ray 
emission in GRBs can establish the {\it mechanism} generating GRBs. Two 
alternative processes have been discussed as the possible dominant 
source of polarization of the prompt $\gamma$-ray emission in GRBs: {\it 
inverse Compton scattering} (ICS) and {\it synchrotron radiation} (SR). 
These are the mechanisms  underlying the prompt $\gamma$-ray 
emission in the  cannonball (CB) model (Dar \& De R\'ujula 2004 and 
references therein) and the `standard' fireball (FB) model 
(see, e.g., Meszaros 2006 and references therein), respectively. In 
the CB model, the ICS of ambient light by highly relativistic jets 
naturally results in a sizable polarization, as predicted 
(Shaviv \& Dar 1995; Dar \& De R\'ujula 2004) years before the first GRB
polarization measurements were made. 

The polarization of 
the prompt $\gamma$-ray emission has been measured in four bright GRBs: 
GRB 021206, GRB 930131, GRB 960924 and GRB 041219a. The first measurement,
made by Coburn and Boggs (2003) with the Reuven Ramaty High 
Energy Solar Spectroscopic Imager (RHESSI) satellite, found a linear 
polarization, $\Pi= (80\pm 20)\%$, of the  $\gamma$-rays of 
GRB 021206.  Subsequent analyses 
by other groups  did not confirm this result at the same level of 
significance (Wiggler et al. 2004; Rutledge \& Fox 2004), so that the 
degree of polarization of GRB 021206 remained uncertain. Later, Willis et 
al.~(2005) have used the BATSE instrument on the Compton Gamma Ray 
Observatory (CGRO) to measure, for two GRBs, the angular distribution of 
$\gamma$-rays back-scattered by the rim of the Earth's atmosphere: 
$35\%\leq \Pi\leq 100\%$ for GRB 930131 and $50\%\leq \Pi\leq 100\%$ for 
GRB 960924. Using coincidence events in the SPI 
(the Spectrometer on the INTEGRAL satellite) and IBIS (the Imager on Board 
the INTEGRAL satellite), 
%similarly to the method used by Coburn \& Boggs (2003) with RHESSI, 
Kalemci et al.~(2006) have recently measured
$\Pi=98\% \pm 33\%$  
for GRB 041219a. Conservatively, 
they could not ``strongly rule out the possibility that 
the measured modulation is dominated by instrumental systematics''.

The polarization naturally
expected for synchrotron radiation (SR) from 
electrons that have been shock-accelerated by chaotic magnetic fields
is  very small. This is unless certain extremely contrived conditions
are met, as first proposed (Eichler \& Levinson 2003; 
Waxman 2003; Granot 2003; Nakar, Piran \& Waxman 2003; 
Granot and Konigl 2003; Lazzati 2006) after a very large 
polarization was first detected by Coburn and Boggs (2003). 
The indications of a nearly  maximal polarization in 4 GRBs 
strongly suggest that there should be a simpler explanation.
The observed polarization 
confirms a simple prediction of the CB model, and further invalidates the 
standard FB model and other SR-based models, as we 
proceed to discuss. In the CB model, the predicted 
linear polarization of the prompt emission in X-ray flashes (XRFs) 
is nearly an order of magnitude below maximal.
Its measurement can be used to test 
the CB model and its unification of GRBs and XRFs.

\section{The polarization of the prompt $\gamma$-ray emission in the CB  
model}

In the CB model (Dar \& De R\'ujula~2000a, 2004), {\it long-duration} GRBs 
and their afterglows (AGs) are produced by bipolar jets of CBs, ejected in 
core-collapse supernova (SN) explosions (Dar \& Plaga~1999).
It is  hypothesized that  an accretion disk is 
produced around the newly formed compact 
object, either by stellar material originally close to the surface of the 
imploding core and left behind by the explosion-generating outgoing shock, 
or by more distant stellar matter falling back after its passage (De 
R\'ujula~1987). As observed in microquasars, each time part of the disk 
falls abruptly onto the compact object, a pair of CBs made of {\it 
ordinary plasma} are emitted with high bulk-motion Lorentz factors, 
$\gamma={\cal{O}}(10^3)$, in opposite directions along the rotation axis, wherefrom matter 
has already fallen onto the compact object, due to lack of rotational 
support. The $\gamma$-rays of a single pulse in a GRB are produced as a CB 
coasts through the SN {\it glory} --the SN light scattered by the SN and pre-SN 
ejecta. The electrons enclosed in the CB Compton up-scatter glory's 
photons to GRB energies. Each pulse of a GRB corresponds to one CB. The 
baryon number, Lorentz factor, and emission time of the individual CBs 
reflect the chaotic accretion process and are not currently predictable, 
but given these parameters (which we extract from the analysis of GRB 
AGs), all properties of the GRB pulses follow (Dar \& De R\'ujula~2004).

Let primed quantities refer to a CB's rest system and unprimed ones
to the SN's rest system. In the CB's system, 
%because of the large value of $\gamma$,
the bulk of the glory's photons
---of energy $E_i={\cal{O}}(1)$ eV in the SN rest frame---
are incident almost in the direction of relative motion,
$\theta_i'={\cal{O}}(1/\gamma)$. Their energy
is $E'_i={\cal{O}}(\gamma\,E_i)\ll m_e\,c^2$,
so that their Compton cross section is in the  low-energy
``Thomson'' limit. 
Let $\theta'$ be the 
angle  at which a photon is 
scattered by a CB's electron\footnote{The Thomson cross section
is $\propto 1+\cos^2(\theta')$, so that
$\theta'\gg\theta'_i\simeq 0$ is an excellent approximation,
but for extremely forward-scattering events that do not result
in observable photons at GRB energies.}. It is  related to the observer's
angle $\theta$ (relative to the CB's direction of motion) by:
\begin{equation}
  \cos\theta' = {\cos\theta-\beta \over 1-\beta\, \cos\theta}\, .
\label{thetaprime}
\end{equation}
The scattering linearly polarizes the outgoing photons in the direction
perpendicular to the scattering plane by an amount
(e.g.~Rybicki \& Lightman 1979):
\begin{equation}
    \Pi(\theta')\approx {1-\cos^2\theta'\over 1+\cos^2\theta'}.
\label{polarcb}
\end{equation}
Substitute Eq.~(\ref{thetaprime}) into Eq.~(\ref{polarcb}) to obtain the
value of the (Lorentz-invariant) linear polarization in the observer's
frame.
In the large-$\gamma$ approximation, the result is
(Shaviv \& Dar 1995, Dar \& De R\'ujula 2004;
see also the recent work\footnote{Some
``sociological'' aspects of these papers have been discussed in De
R\'ujula (2003).} of Lazzati et al.~2004 and Lazzati~2006): 
\begin{equation}
\Pi(\theta,\gamma)\approx {2\;\theta^2\,\gamma^2\over 
1+\theta^4\,\gamma^4},
\label{polSN}
\end{equation}
which, for the most probable viewing angles,
$\theta\approx 1/\gamma$, is of ${\cal{O}}(100\%)$ .
This result is easy to understand: For $\gamma\gg 1$, photons viewed at 
$\theta =1/\gamma$, i.e. at $\cos(\theta)\approx \beta$,
were scattered at $\theta'=90^{\rm o}$, according to Eq.~(\ref{thetaprime}),
acquiring a nearly total polarization, according to
Eq.~(\ref{polarcb}). 
In the CB model, GRBs with extremely large equivalent 
isotropic luminosities are viewed almost on-axis. For  them,   
$\theta^2 \gamma^2 \ll 1$, and their polarization is 
$\Pi\approx 2\, \gamma^2\, \theta^2\ll 1$.

\section{The polarization of prompt $\gamma$-ray emission in the FB models}

Synchrotron radiation from a power-law distribution of electrons 
$dn_e/dE\sim E^{-p}$ in a {\it constant, uni-directional} 
magnetic field can produce a 
large polarization (e.g., Ginzburg and Syrovatski, 1969; Rybicki and 
Lightman 1979; Longair 1994), $\Pi=(p+1)/(p+7/3)\approx 70\%$,
for the canonical power-law index, $p\approx 2.2$. But
collisionless shock-acceleration requires highly 
disordered and time varying magnetic fields (for a recent review see, 
e.g., Zhang \& Meszaros 2003, for a dissenting view, 
see Lyutikov, Pariev \& Blandford 2003). Only under very contrived 
circumstances, which 
should be the exception and not the rule ---such as geometrical 
coincidences, and unnaturally ordered magnetic fields--- can 
collisionless shocks produce a large polarization. In 
our opinion, this is what various articles (Eichler \& Levinson, 2003; 
Waxman, 2003; Granot 2003; Nakar, Piran \& Waxman, 2003; 
Granot and Konigl 2003; Lazzati 2006) on the subject show, 
although it is not what their authors conclude.

In fireball models, the origin of the prompt 
and  afterglow emissions is synchrotron radiation
from mergers (of two fireshells), and from
the collision of the ensemble of shells
with the interstellar medium (ISM),  respectively. No fundamental 
difference between these collisions is assumed (even though the center 
of mass energy is larger in the collision with the ISM, implying 
a larger energy release in the afterglow, in blatant
contradiction with the observations).
Since a linear polarization 
is Lorentz invariant, both types of collision
should produce a low polarization. Indeed,
linear polarizations of the order of a few percent were proposed to 
arise from causally-connected magnetic patches (e.g.~Gruzinov \& 
Waxman 1999), from homogeneous conical jets (Gruzinov 1999; 
Ghisellini \& Lazzati 1999; Sari 1999) and from structured jets viewed 
off-axis (Rossi, Lazzati \& Rees 2002).  
Small linear polarizations of the optical afterglow have 
been detected (see e.g.~Covino et al.~2005 for a review).

One may argue that, due to relativistic beaming, only  a 
small area of a  fireshell  merger is visible, 
and that in such a small patch the magnetic field may 
be aligned along a single direction. Later, during the afterglow phase, 
when a larger area becomes visible due to the decreasing 
Lorentz factor of the relativistic ejecta, the visible area contains many 
patches with a magnetic field aligned in a random direction, yielding a 
small total polarization.  However, even if the magnetic field were 
aligned 
in one direction in the visible patch of a fireshell merger, 
it is unlikely to be aligned in the same direction in the different 
fireshell mergers that produce the different peaks of a multi-peak GRB. 
The time-integrated polarization of a multi-peak GRB, such as
GRB 021206, should average to a very small number.

Finally, in shock-acceleration models which invoke synchrotron 
self-absorption to explain the low-energy spectral shape of GRBs, the 
linear polarization of self-absorbed SR is small (Ginzburg and Syrovatski, 
1969; Longair 1994) $\Pi\!=\!3/(6p+3)\!<\! 20\%$.
For a Band spectrum,  most of the photons have an energy below the peak 
energy.  If the peak-energy feature results from self-absorption of the 
lower-energy SR, the total polarization cannot exceed  $\sim\!20$\%. 

It may not come as a surprise that SR fails to 
explain a large polarization of a GRB's  $\gamma$ rays.
It has been known for long  that SR fails in accommodating
the observed spectrum (Ghirlanda et al.~2004) 
and that the a-posteriori attempts
to explain the ratio of prompt and afterglow
total energies (an `energy crisis', Piran 1999) are not convincing.

\section{The transition from GRBs to XRFs}

The Doppler factor boosting the energy of radiation in the CB's
rest system to the SN's rest system is:
\begin{equation}
\delta \equiv {1\over\gamma\,(1-\beta\, cos\theta)}
                       \simeq  {2\, \gamma
                       \over 1+\gamma^2\, \theta^2}\; ,
\label{delta}
\end{equation} 
where the approximation is excellent 
for  $\theta\!\ll\! 1$ and  $\gamma \!\gg\! 1$. In the CB model, 
XRFs and GRBs are the same phenomenon,
viewed from different observer's angle $\theta$
(Dar \& De R\'ujula 2000a, 2004; Dado et al.~2004).
For GRBs, $\theta\,\gamma\sim 1$, while for XRFs it is larger
(Dar \& De R\'ujula 2000a,  2004; Dado et al.~2004).  
The various GRB and XRF observables scale as powers of 
$\gamma$ and $\delta$. For example, the typical prompt photon 
energy is $E_\gamma\propto \gamma\,\delta$, the
spherical equivalent energy is $E_\gamma^{\rm iso}\propto \delta^3$, 
and the peak isotropic luminosity is $L_p^{\rm iso}\propto \delta^4$,
implying various correlations between these observables
(Dar \& De R\'ujula 2000b, Dado et al.~2006, 2007) and the gradual
evolution from the `hard' GRBs to the `softer' XRFs.

For XRFs, Eq.~(\ref{polSN}) implies a smaller polarization, 
$\Pi\approx 2/\gamma^2\theta^2\ll 1$, than for GRBs.
For typical XRFs, $\theta\,\gamma\!>\! 4$ and  $\Pi\, \lsim\, 12\%$,
as shown in Fig.~1.
Thus, polarization measurements of the prompt emission in XRFs 
can provide another crucial test of the CB model and its unification of
GRBs and XRFs.

\section{Conclusions}

A large linear polarization of the prompt $\gamma$-ray emission in 
ordinary GRBs, such as that indicated by data on
GRBs 021206,  930131,  960924 
and  041219a, provides strong support for the CB model, wherein inverse 
Compton scattering of ambient light by highly relativistic CBs is the 
dominant production mechanism of the prompt $\gamma$-ray emission. These 
observations are inconsistent with the expectation from the standard 
fireball model, wherein synchrotron radiation from shock-accelerated
 electrons is the assumed production mechanism. The CB model 
also predicts a  small polarization of the prompt emission in XRFs, 
and in GRBs lying at the upper end of the distributions
of peak and equivalent isotropic luminosities. 
Further polarization measurements of XRFs and 
GRBs will provide a decisive test of the CB model unification of 
these phenomena, as well as extra support to inverse Compton scattering
as the dominant source of their prompt radiations.

\begin{figure}[]
\epsfig{file=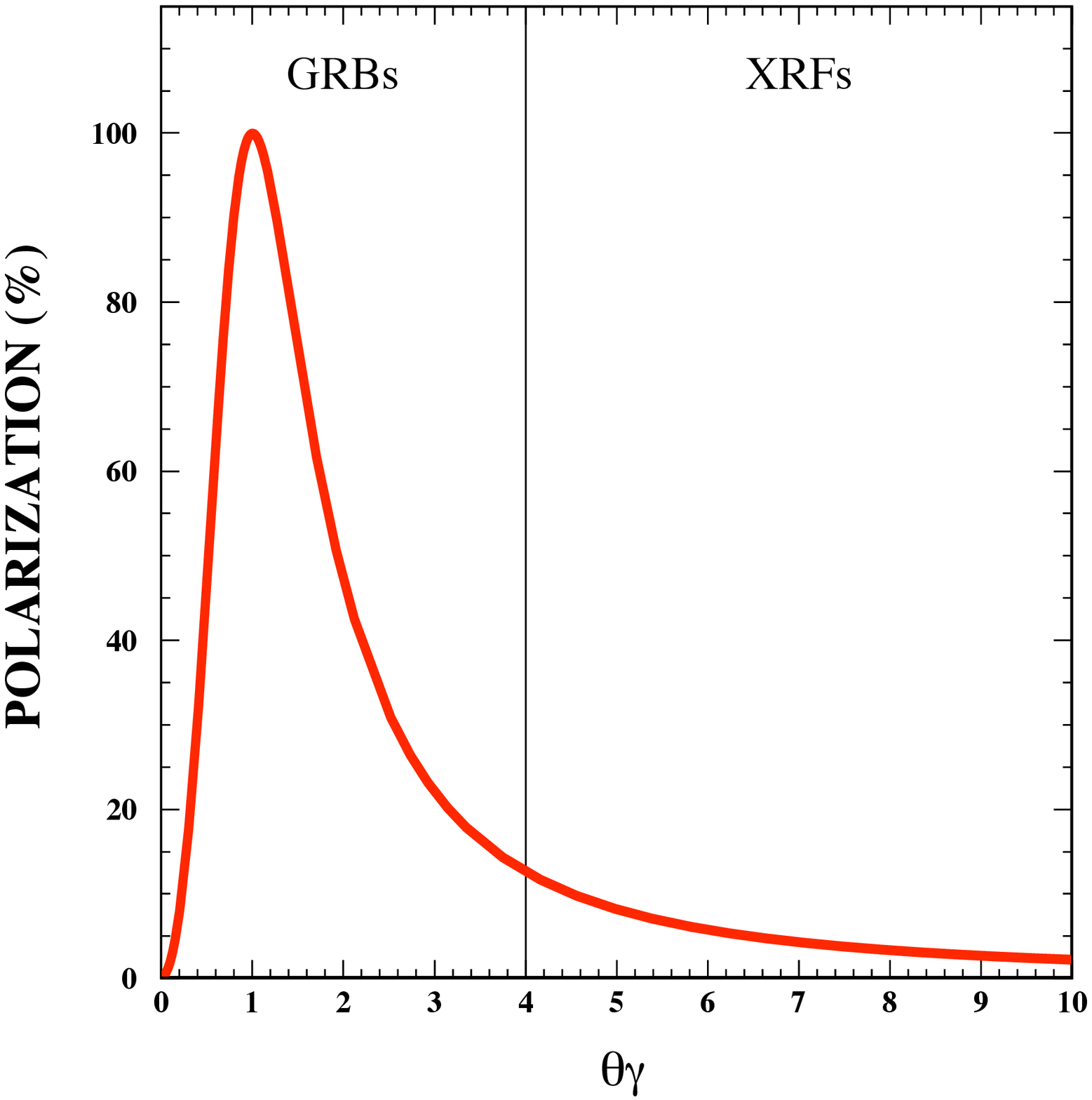,width=16cm}
\vspace*{8pt}
\caption{The linear polarization predicted by the CB model for the prompt 
emission in GRBs and XRFs as function of $\theta\gamma$.
The vertical line
$\theta\gamma=4$ roughly indicates the transition from GRBs to XRFs.} 
\label{f1}
\end{figure}

\end{document}